# American Put Option pricing using Least squares Monte Carlo method under Bakshi, Cao and Chen Model Framework (1997) and comparison to alternative regression techniques in Monte Carlo

Anurag Sodhi
University of North Carolina at Charlotte

**Abstract**

This paper explores alternative regression techniques in pricing American put options and compares to the least-squares method (LSM) in Monte Carlo implemented by Longstaff-Schwartz, 2001 which uses least squares to estimate the conditional expected payoff to the option holder from continuation. The pricing is done under general model framework of Bakshi, Cao and Chen 1997 which incorporates, stochastic volatility, stochastic interest rate and jumps. Alternative regression techniques used are Artificial Neural Network (ANN) and Gradient Boosted Machine (GBM) Trees. Model calibration is done on American put options on SPY using these three techniques and results are compared on out of sample data.

May 8, 2018

I would like to thank to my professor, Dr. Steven Clark for wonderful guidance throughout the Advanced Derivatives course that he taught.

# Contents

# 1. Introduction

American options, compared to European counterparts, have an additional flexibility of exercising at any time before maturity, which adds to complexity in pricing. Monte Carlo simulation is a rather flexible valuation approach which is applicable to almost any feature a financial product can exhibit: American and Bermudan exercise, Asian and lookback features (i.e. path dependency).

For pricing American options, nested Monte Carlo can be used, but is very computationally expensive. In 2001, Longstaff-Schwartz proposed least-squares method (LSM) in Monte Carlo which uses least squares to estimate the conditional expected payoff to the option holder from continuation. Although this reduced the computational time drastically, it introduces approximation error from the least squares regression. In this paper we explore alternative regression techniques like Artificial Neural Networks (ANN) and Gradient Boosted Machine (GBM) Trees and compare the results with LSM method to check whether we can improve the quality of the regression.

The paper is structured as follow. Section 2 provides general introduction to Least squares method implemented by Longstaff-Schwartz. It also reviews the definitions and explanation of the approaches used in the study. Section 3 contains a description for data and in Section 4, I will describe in detail the analysis and approaches were used to calculate the results. Section 5 concludes and outlines the direction for future research.

# 2. Least-square Monte Carlo valuation for American Options

Price of American Put Option depends on the optimal stopping time $\tau$, and is given by:

$$V_0 = \sup_{\tau \in [0,T]} \mathbf{E}^Q_0 [B_0(\tau) \, h_\tau(S_\tau)]$$

with $V_0$ being the present value of the American derivative, $\tau$ an F-adapted stopping time, $T$ the date of maturity, $B_0(\tau)$ the discount factor appropriate for stopping time $\tau$, $h_\tau$ a nonnegative, $\tau$-measurable payoff function and $S_\tau$ the index level process stopped at $t = \tau$. The expectation is again taken under the risk-neutral measure Q.

The continuation value at time t using the Markov property of $S_t$, which is used to calculate the option value at time t as $V_t(s) = \max[h_t(s), C_t(s)]$ is given by:

$$C_t(s) = \mathbf{E}^Q_t [e^{-r\Delta t} V_{t+\Delta t}(S_{t+\Delta t}) | S_t = s]$$

The LSM method utilizes the fact that we know the simulated continuation value at time t for $i^{th}$ path but cant use it directly, otherwise it would be perfect foresight: $Y_{t,i} \equiv e^{-r\Delta t} V_{t+\Delta t, i}$

Longstaff-Schwartz (2001) proposed to regress the these $Y_{t,i}$ against the simulated index levels $S_{t,i}$ with D basis functions b;

$$\hat{C}_{t,i} = \sum_{d=1}^{D} a_{d,t} * b_d(S_{t,i})$$

The optimal regression parameters $\alpha^*_{d,t}$ are the result of the minimization:

$$\min_{\alpha_{1,t},\dots,\alpha_{D,t}} \frac{1}{I} \sum_{i=1}^{I} \left( Y_{t,i} - \sum_{d=1}^{D} \alpha_{d,t} \cdot b_d(S_{t,i}) \right)^2$$

In the next section we explore how ANN and GBM can be used instead of D basis functions for regression. We also include variance reduction techniques: antithetic values, moment matching and control variates for the simulation.

## 3. Universal Approximation Theorem

For any function $f$, we can define F(x) as an approximate realization of $f$, for some constants $v_i, b_i$, vector $\omega i$ and $\varphi$ a nonconstant, bounded, and monotonically-increasing continuous function:

$$F(x) = \sum_{i=1}^{N} v_i \varphi(\omega_i^T x + b_i)$$

In the following sub-section we see how this is similar to a one layer ANN and GBM regression.

### i. Artificial Neural Network (ANN)

Artificial neural networks are computational models inspired by biological neural networks, consisting of:
- Input layer
- One of more hidden layers
- Output layer

ANN is parametrized by number of neurons, weights, biases and activation function of each layer. This is shown in the Figure 1 and 2.

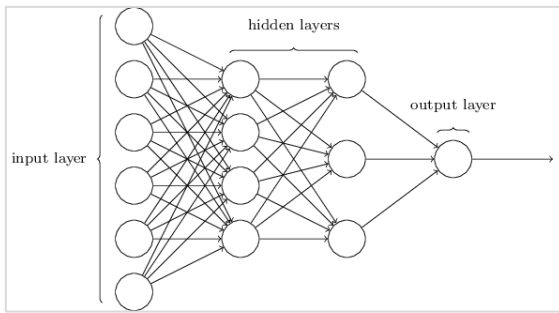
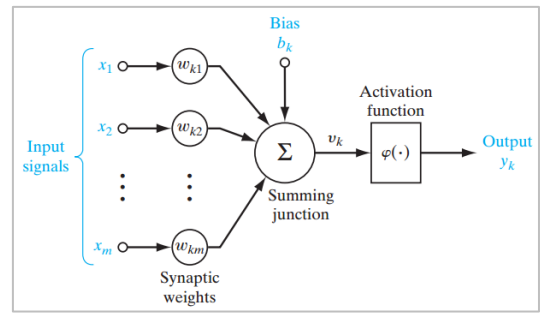

*Figure 1*  *Figure 2*

In Figure 2, we see neurons, each characterized by a weight. For each input $x_i$ in the neuron k there is a synaptic weight $w_{kj}$ to multiply it. A linear combiner or adder for summing the weighted input signals and bias $b_k$. An activation function $\varphi$ (.) for limiting the amplitude of the neuron's output.

This is essentially what Universal Approximation theorem says and therefore any function can be approximated by a one hidden layer ANN. More formal proof can be found in Balázs Csanád Csáji (2001). "Approximation with Artificial Neural Network". However, in practice we might require more than one layer to reach the convergence through gradient descent early. Figure 3 below shows an example where a 4 layer ANN (50 neurons each) is used to approximate a function.

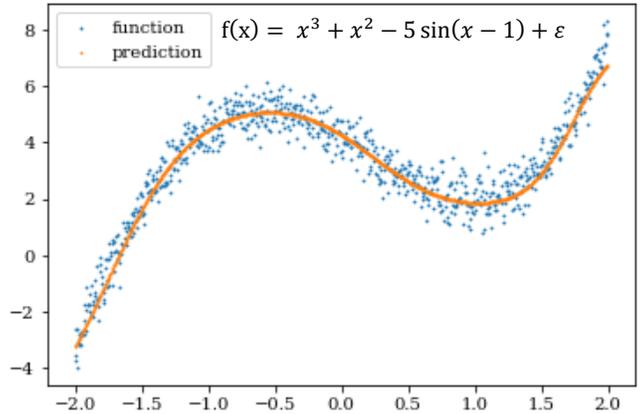

*Figure 3: ANN approximating a function*

### ii.  Gradient Boosting Machine (GBM)

GBM is an ensemble of decision trees. It uses a simple regression model to start but subsequent models predict the error residual of the previous predictions. It trains learners sequentially by focusing on predictions that other trees got wrong. Overall prediction given by a weighted sum of the collection. More details on GBM can be found in Friedman, J. H. (1999). "Greedy Function Approximation: A Gradient Boosting Machine". Figure 1 shows how GBM iterates during the learning and Figure 2 gives an example for function approximation.

[2]

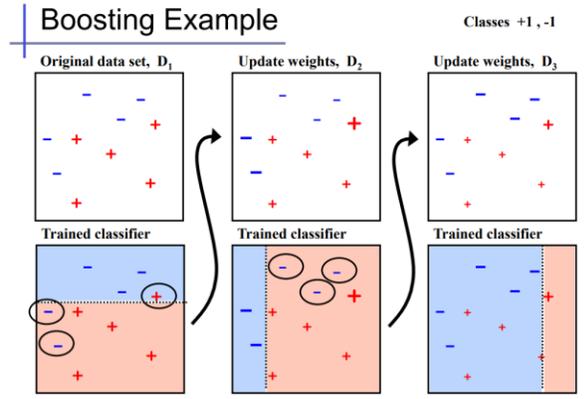
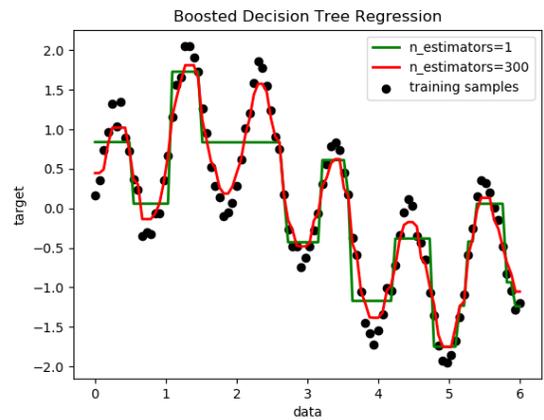

*Figure 4*                         *Figure 5*

### iii.  Quick comparison of function approximation: LSM, GBM and ANN

The figure below provides a visual comparison of the fit in the three approaches. LSM here uses 5 basis functions (polynomials up to degree 5). The x-axis show simulated stock price at a step in time and the y-axis show the discounted option value from the next step. From the Figure 6 we can see polynomial might not always provide the best fit. The fit in GBM and ANN is much better visually.

---

[2] : http://sli.ics.uci.edu/Classes/2012F-273a?action=download&upname=10-ensembles.pdf

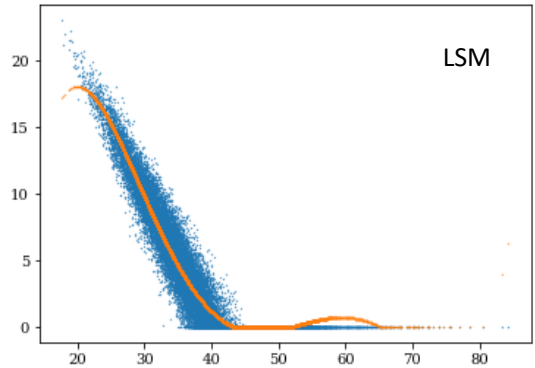

*Figure 6*

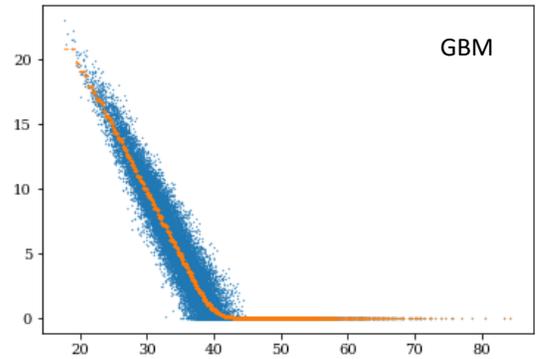

*Figure 7*

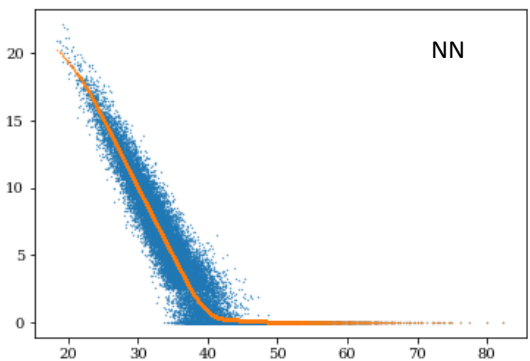

*Figure 8*

## 4. Comparison of an example American put pricing for LSM, GBM and ANN

Here we compare pricing of an example American put option with the following details:

S0 = 36, K = 40, T = 1.0, r = 0.06, sigma = 0.2

Binomial Tree Price with 500 steps: **4.486**

Monte Carlo simulation was performed with 20 time-intervals and 25000 paths. Table 1 shows the results.

|  | **LSM** | **GBM** | **ANN** |
|---|---|---|---|
| **Option Value** | 4.472 | 4.509 | 4.474 |
| **Total Time (sec)** | 0.41 | 1.72 | 13.88 |
| **Parameter Details** | Number of Basis functions (D) = 5 | Estimators = 50 | 4 hidden layers of 20 neurons each, Batch Size =256 |

*Table 1*

# 5. Pricing and Calibration on real data

### i. Modeling Methodology and Framework

Bakshi, Cao and Chen Model (1997), BCC97 was used as the underlying framework as it incorporates stochastic interest rates, stochastic volatility and jumps. The dynamics of index S under risk neutral measure are as follows

$$dS_t = (r_t - r_J)S_t\, dt + \sqrt{v_t}S_t dZ_{t1} + J_t S_t dN_t$$

$$dv_t = k_v(\theta_v - v_t)dt + \sigma_v \sqrt{v_t} dZ_{t2}$$

The interest used in the BCC97 are also stochastic and follow CIR model:

$$dr_t = k_r(\theta_r - r_t)dt + \sigma_r \sqrt{r_t} dZ_{t3}$$

Monte Carlo approach as described in Section 2 was used as the modeling methodology. Three models were built and calibrated, one for each type of regression used: LSM, GBM, ANN. Least squares estimation with basis function as described in Section 2 was replaced by GBM and ANN method as described in Section 3 for the other two models. Monte Carlo simulation requires discretization for simulating the index paths, which was done according to Yves Hilpisch (2015) "Derivatives Analytics with Python":

$$S_t = S_s \left( e^{(\bar{r}_t - r_J - v_t/2)\Delta t + \sqrt{v_t}\sqrt{\Delta t} z_t^1} + \left(e^{\mu_J + \delta^2 z_t^4} - 1\right) y_t \right)$$

$$\tilde{v}_t = \tilde{v}_s + \kappa_v(\theta_v - \tilde{v}_s^+)\Delta t + \sigma_v \sqrt{\tilde{v}_s^+}\sqrt{\Delta t} z_t^2$$
$$v_t = \tilde{v}_t^+$$
$$\tilde{r}_t = \tilde{r}_s + \kappa_r(\theta_r - \tilde{r}_s^+)\Delta t + \sigma_r \sqrt{\tilde{r}_s^+}\sqrt{\Delta t} z_t^3$$
$$r_t = \tilde{r}_t^+$$
$$\bar{r}_t \equiv (r_t + r_s)/2$$

- $S_t$ index level at date $t$
- $r_t$ risk-less short rate at date $t$
- $r_J \equiv \lambda \cdot (e^{\mu_J + (\delta^2)/2} - 1)$ drift correction for jump
- $v_t$ variance at date $t$
- $\kappa_v$ speed of adjustment of $v_t$
- $\theta_v$ the long-term average of the variance, $\theta_r$ the long-term average of the short rate
- $\sigma_v$ volatility coefficient of the index's variance, $\sigma_r$ volatility coefficient of the short rate
- $Z_t^n$ standard Brownian motions
- $N_t$ Poisson process with intensity $\lambda$
- $J_t$ jump at date $t$ with distribution: $\log(1 + J_t) \approx \mathbf{N}(\log(1 + \mu_J) - \delta^2/2, \delta^2)$

## ii. Calibration Methodology

Model calibration was done in 4 parts:

1. Calibrate CIR model for stochastic interest rates from zero coupon prices to get $\kappa r, \theta r, \sigma r$ (local optimization)

2. Calibrate Heston '93 (Stochastic interest) model incorporating stochastic interest rate calibrated above to get $\kappa v, \theta v, \sigma v, \rho, v0$ (grid search + local optimization)

3. Calibrate Jump part of the BCC97 model using short maturity options and get $\lambda, \mu, \delta$ (grid search + local optimization)

4. Run local optimization over the complete BCC97 model to get new $\kappa v, \theta v, \sigma v, \rho, v0, \lambda, \mu, \delta$

## iii. Data gathering and preparation

Data gathering:

- Data on American put options on SPY (SPDR S&P 500 ETF) from OptionMetrics
- Zero Coupon bond data
- SPY index price from yahoo finance

Data Preparation:

- Chose September 5th, 2017 options and zero-coupon bond data for calibrating.
- Kept option in 80% to 120% moneyness and with maturities ranging from ~1 to ~6 weeks (OptionMetrics data had the max. maturity of around 6 weeks)
- Removed options low trading volume less than 50 and prices less than 10 cents
- Total of 67 options

Out of sample data preparation:

- Options for next week in the same month - Total 166 options
- Options for a week in next month – Total 117 options

## iv. Calibration Results

For the interest rate, the following parameters were obtained:

$$\kappa r, \theta r, \sigma r = 0.123, 0.066, 0.001$$

The table below compares the parameters in the BCC 97, for LSM, GBM and ANN models after calibration:

| Model | $\kappa v$ | $\theta v$ | $\sigma v$ | $\rho$ | $v0$ | $\lambda$ | $\mu$ | $\delta$ |
|---|---|---|---|---|---|---|---|---|
| LSM | 20.850 | 0.012 | 0.712 | -0.984 | 0.002 | 0.0001 | -0.378 | 0.0005 |
| GBM | 22.568 | 0.011 | 0.718 | -0.997 | 0.001 | 0.0001 | -0.361 | 0.0004 |
| ANN | 15.141 | 0.012 | 0.481 | -0.967 | 0.003 | 0.0001 | -0.377 | 0.0004 |

*Table 2*

### v. In-sample and Out-of-sample Pricing Results

The table 3 below shows Mean Squared error for (MSE) for LSM, GBM, ANN.

| MSE | LSM | GBM | ANN |
| --- | --- | --- | --- |
| **In-Sample** | 1.018 | 0.989 | 0.999 |
| **Out-of-Sample (1 week after calibration)** | 0.932 | 0.868 | 0.857 |
| **Out-of-Sample (1 month after calibration)** | 0.809 | 0.681 | 0.649 |

*Table 3*

Table 4 below provides a bit details view of MSE by maturity and moneyness. In the table below, Near the money (NTM) options are grouped when strike is around 2% the index. Out of money (OTM) are those grouped options when stock is more than 2% greater than strike and rest are defined in a group called In the money (ITM). Short Maturity is defined as ~1 to ~3 weeks maturity and Mid Maturity is defined as ~3 to ~6 weeks maturity.

| MSE | | LSM | | | GBM | | | ANN | | |
| --- | --- | --- | --- | --- | --- | --- | --- | --- | --- | --- |
| | Maturity / Moneyness | ITM | NTM | OTM | ITM | NTM | OTM | ITM | NTM | OTM |
| **In-Sample** | Short | 9.48 | 0.79 | 0.02 | 9.48 | 0.75 | 0.03 | 9.48 | 0.77 | 0.03 |
| | Mid | 3.94 | 0.22 | 0.07 | 3.80 | 0.17 | 0.09 | 3.80 | 0.16 | 0.13 |
| **Out-of-Sample (1 week after calibration)** | Short | 5.34 | 0.99 | 0.01 | 5.33 | 0.92 | 0.01 | 5.34 | 0.94 | 0.02 |
| | Mid | 3.97 | 0.71 | 0.03 | 3.77 | 0.51 | 0.04 | 3.77 | 0.40 | 0.06 |
| **Out-of-Sample (1 month after calibration)** | Short | 5.42 | 0.92 | 0.01 | 5.36 | 0.81 | 0.01 | 5.36 | 0.84 | 0.01 |
| | Mid | -- | 1.31 | 0.02 | -- | 0.99 | 0.01 | -- | 0.82 | 0.01 |

The results show that ANN performs the best followed by GBM and then the LSM.

## 6. Conclusion and Next steps

Although both GBM and ANN had better results than LSM, if we include time taken in pricing then LSM is much better, GBM is a close second but ANN takes a lot of time. With newer and newer hardware and GPU optimized deep learning packages like Tensorflow, Theano we might see a significant reduction in ANN computation time. The choice of finer grid in calibration and more optimization steps affects the MSE a lot since the objective function can get stuck in local minima.

For next steps we could re-run the calibration with MSE of relative price difference instead of absolute price difference. We can also explore using variance gamma process instead of Merton's jump diffusion process in the BCC 97 framework for a more parsimonious model as it uses less parameters.

# 7. References

1. Gurdip Bakshi, Charles Cao, and Zhiwu Chen (1997). "Empirical Performance of Alternative Option Pricing Models"
2. Francis A. Longstaff and Eduardo S. Schwartz (2001). "Valuing American Options by Simulation: A Simple Least-Squares Approach"
3. Balázs Csanád Csáji (2001). "Approximation with Artificial Neural Network"
4. Friedman, J. H. (1999). "Greedy Function Approximation: A Gradient Boosting Machine"
5. Cox, J.C., J.E. Ingersoll and S.A. Ross (1985). "A Theory of the Term Structure of Interest Rates"
6. Yves Hilpisch (2015) "Derivatives Analytics with Python"
7. Python's scikit-learn package's MLP (Multi-layer Perceptron) Regression used for ANN: http://scikit-learn.org/stable/modules/neural_networks_supervised.html#regression
8. XGBoost package used for GBM: http://xgboost.readthedocs.io/en/latest/python/index.html

# 8. Code for this project

Anurag_derivatives_project.zip

The above zip file doesn't contain the data. The code and the data can also be downloaded from:

https://www.dropbox.com/sh/4g57fh1xq7h77h0/AADIm99xOafk-6WfjcKxdbNEa?dl=0